\documentclass[twocolumn,prl,aps,epsf,showpacs]{revtex4-1}

\usepackage{graphicx}
\usepackage{dcolumn}
\usepackage{bm}

\begin{document}

\title{Colossal magnetoresistance in an ultra-clean  weakly interacting 2D Fermi liquid}

\author{Xiaoqing Zhou$^{1}$, B.A. Piot$^{2}$, M. Bonin$^{1}$, L.W. Engel$^{3}$, S. Das Sarma$^{4}$, G. Gervais$^{1}$, L.N. Pfeiffer$^{5}$ and K.W. West$^{5}$}

\affiliation{$^{1}$ Department of Physics, McGill University,
Montreal, H3A 2T8, CANADA}

\affiliation{$^{2}$ Laboratoire National des Champs MagnŽtiques Intenses, Centre National de la Recherche
Scientifique, 25 Avenue des Martyrs, F-38042 Grenoble, France}

\affiliation{$^{3}$ National High Magnetic Field Laboratory, Tallahassee, FL 32310, USA}

\affiliation{$^{4}$ Condensed Matter Theory Center, Department of
Physics, University of Maryland, College Park, MD 20742 USA}

\affiliation{$^{5}$ Department of Electrical Engineering, Princeton University, Princeton NJ 08544 USA}

\date{\today }

\begin{abstract}

We report the observation of a new phenomenon of colossal magnetoresistance in a 40 nm wide GaAs quantum well in the presence of an external magnetic field applied parallel to the high-mobility 2D electron layer. In a strong  magnetic field, the magnetoresistance
is observed to increase by a factor of $\sim$300 from 0 to $45T$ without the system undergoing any metal-insulator transition.  We discuss how this colossal magnetoresistance effect cannot be attributed to the spin degree-of-freedom or localization physics, but most likely emanates from strong magneto-orbital coupling between the  two-dimensional electron gas and the magnetic field. Our observation is consistent with a field-induced 2D-to-3D transition in the confined electronic system.

\end{abstract}
\pacs{73.43.Qt, 75.47.De, 75.47.Gk } \maketitle

We report in this Letter the experimental discovery of a Colossal Magnetoresistance (CMR) effect in an ultra-clean and (relatively) high-density two-dimensional  (2D) electron system. The 2D resistivity of a 40 nm wide modulation-doped GaAs quantum well  (with a carrier density $n\simeq 10^{11}$ $cm^{-2}$ and an ultra-high mobility $\mu\simeq 10^{7}$ $cm^{2}/V\cdot s$) is found to increase by a factor of 10, 30,  and 300 respectively, for an applied magnetic field of $\sim$8T, 15T,  and 45T oriented {\it parallel} to the 2D plane,  {\it i.e.} in zero perpendicular field.  Our observed CMR  phenomenon is unrelated to the earlier observations of 2D parallel-field magnetoresistance  \cite{Simonian97,Pudalov, Okamoto99,Yoon00,PadpadakisANISO00,Tutuchole01,Vitkalov,Gao02,Zhu,Gao06,Lai,Lu,Piot2009}   where either spin-polarization or disorder-related localization (or both) play decisive roles. Our high-mobility ultra-pure 2D system remains metallic, with metallicity parameter $k_{F}\lambda \gg 1$ ($\lambda$ is the transport mean free path) over the whole $B_{\parallel}=0-45T$ applied parallel field range, implying that strong localization effects associated with 2D metal-insulator transition (MIT)  play no role in our observation. This contrasts with many recent experiments  \cite{Simonian97,Pudalov, Okamoto99,Yoon00,PadpadakisANISO00,Tutuchole01,Vitkalov,Gao02,Zhu,Lai,Gao06,Lu,Piot2009} where $k_{F}\lambda \lesssim 1$ in the high-field regime. Furthermore, due to the relatively high electronic density of our system, the field-induced 
carrier spin-polarization is extremely small, {\it i.e.} $\frac{E_z}{ E_{F}} \ll1$ where $E_{z}=g^{*}\mu_{B} B_{\parallel}$ is the Zeeman splitting and $E_{F}$ the Fermi energy, even at  our highest applied field; in fact, a full spin-polarization of our 2D system would require a field $B_{\parallel}> 100T$. To the best of our knowledge, this observation of a two and a half orders of magnitude CMR  effect is by far the strongest magnetoresistance  ever reported in a metallic 2D electron system without the manifestation of a 2D MIT.  We believe that our  observed CMR arises from the parallel-field induced 2D-to-3D transition, although we cannot completely rule out the interesting possibility of an unknown mechanism playing a role here.

Our main experimental findings, presented in Fig.1, can  be summarized as follows: {\it i)} The 2D resistivity increases with increasing applied parallel field  in our 40 nm wide GaAs quantum well with the net increase
being very large, and almost a factor of 300 (50) at 45 (15)T. {\it ii)} The magnetoresistance is much weaker in a 30 nm sample. {\it iii)} The system remains metallic throughout; in fact, the metallicity is stronger at higher magnetic fields, {\it i.e.} the coefficient $\frac{\partial \rho}{\partial T}>0$ is  larger at high fields.  {\it iv)}  Even in the most resistive situation, {\it i.e.} at the highest applied fields,  the measured resistivity remains well below the Ioffe-Regel limit with $k_{F}\lambda \simeq 4000-10$ in the $0-45T$ applied field range, thus keeping our system deep in the effective metallic regime throughout the CMR phenomenon. {\it v)}  The spin polarization  remains small in our experiment, less than $\sim 10\%$. Collectively, these observations are unprecedented and intriguing. The observed   lack of any obvious insulating phase as well as the high mobility
and density of our sample point to our observed  CMR phenomenon being of  magneto-orbital origin, where the parallel field non-perturbatively couples the 2D dynamics of the system with transverse dynamics  ({\it i.e.} normal to the 2D plane) leading to a novel 2D-3D transition producing the CMR effect. Since the system most likely remains a relatively weakly interacting Fermi liquid with a dimensionless interaction parameter $r_{s}\simeq 1.8$ in 2D and $r_{s}\simeq 1$ in 3D for our GaAs quantum well system, and with the disorder parameter $(k_{F}\lambda)^{-1}$ and spin-polarization parameter $(\frac{E_z}{ E_{F}})$ both being very small, it is reasonable to assume that interaction,  localization or magnetization phenomena  are not responsible for the observed CMR effect, in contrast to the other reported 2D parallel field experiments  \cite{Simonian97,Pudalov, Okamoto99,Yoon00,PadpadakisANISO00,Tutuchole01,Vitkalov,Gao02,Zhu,Lai,Gao06,Lu,Piot2009}. 

On the other hand, the width of our quantum well being relatively large ($d=40 $ $nm$), typically much larger than the magnetic length $l_{{\parallel}}\equiv \sqrt{\frac{\hbar c}{eB_{\parallel}}  }$ associated  with the applied parallel magnetic field ($l_{{\parallel}}\simeq \frac{26nm}{\sqrt{B_{\parallel}}} $ $nm$ where $B_{\parallel}$ is in Tesla), leads us to believe that the magneto-orbital coupling \cite{DasSarma2000} induced by the parallel field is producing the CMR effect in our system. This is further corroborated by our measurements on a sample with a quantum well width $d=30$ $nm$ where the CMR, while still being very large, is much less than in the 40 nm sample (see Fig.1). This indicates  that the well width may well be the decisive parameter leading to the CMR effect reported in this work.

The
transport measurements are performed in a sample of  rectangular shape (with
a long-to-short axis ratio $\sim$ 3:1) using a standard low-frequency lock-in technique at a low excitation current, $I=100$ $nA$. Except where noted, all measurements are performed with the current path defined by the contacts set at the edge of the long axis of the rectangle perpendicular to the in-plane magnetic field. The 2D
electron plane is aligned parallel to the magnetic field direction by using an {\it in situ} rotation stage to minimize the Hall voltage $V_{H}=R_{xy}I$ at the highest magnetic field used. In a non-ideal Hall bar or Van der Pauw geometry, when measuring $R_{xy}$, there may be a small  $R_{xx}$
mixing into the measurement of $R_{xy}$, so as a consequence the measured Hall voltage may not necessarily vanish even when   $\theta =90^{o}$ and $B_{\perp}\equiv 0$.  To overcome this, we have performed a systematic study
where we have measured the  longitudinal and Hall resistances on a fine scale in the range $\theta \in[89.6^{o},90.4^{o}]$, up to 12T magnetic field, using a second sample whose Hall voltage allowed a more precise determination of the angle at which the perpendicular field vanishes. In doing so, we have verified that the main observation reported here is not due to a slight misalignment of the magnetic field with the 2D plane. 

The resistance $R_{xx}$ (left axis) and the normalized resistivity $\rho(B_{\parallel})/\rho(0)$ (right axis) as a function of the parallel magnetic field is shown in the panel (a) of  Fig.1, at temperatures $T\simeq 0.18$ K (solid red), 0.55 K (dashed blue) and  1.05 K (dotted purple).  Expanded plots around $8T$ and $12T$ regions are shown in panels (b) and (c). An increasing monotonic magnetoresistance is observed, enhanced by a  factor of  $\sim$40 at $18T$ magnetic field. This colossal magnetoresistance  effect  shows no saturation in very large fields, manifesting an increase by a factor of $\sim$300 at $45T$ at a temperature of $0.51K$, as shown in the inset of Fig.1 (for this data, the current was chosen to be parallel to the in-plane magnetic field).  As a comparison, we also show in Fig.1 the CMR effect observed at 20 mK in a 30 nm wide sample with an electron density  $n\simeq 3\times10^{11}$ $cm^{-2}$ and with a very high mobility,  $\mu \simeq 2
\times 10^7$ $cm^2/(V\cdot s)$. Albeit considerably weaker, the CMR effect in the 30 nm wide quantum well sample is observed to increase the magnetoresistance by a factor of $\sim$10. The more
dramatic CMR effect observed in the 40 nm well is, to our knowledge, by far the strongest ever reported in any metallic 2D system. 

\begin{figure}[tbp]
\includegraphics[width=1.1\linewidth,angle=0,clip]{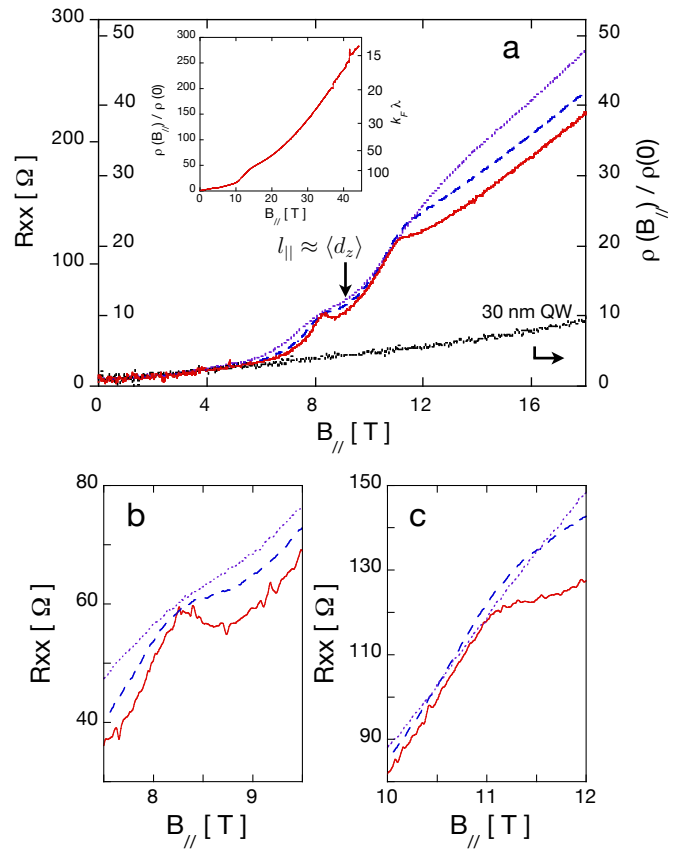}
\caption{a) Longitudinal resistance  $R_{xx}$ as a
function of the parallel magnetic field $B_{\parallel}$ 
at temperatures  $T=0.18K$, (solid red), $0.55K$ (dashed blue)  and
$1.05K$ (dotted purple). The increase in resistivity $\rho(B_{\parallel})/\rho(0)$
is shown on the right scale, where $\rho(0)$ is the $T=0.18K$ zero-field
resistivity. The black line shows the CMR effect at $T=20 mK$ measured in a ultra-high mobility 30 nm wide quantum well, shown here plotted on the right y-axis,  for comparison. Inset: increase in resistance for fields up to $45T$  at $T=0.51K$. The right y-axis gives the value for the metallicity parameter $k_{f}\lambda$ estimated from $\rho(B_{\parallel})$.
b,c) Expanded plot in the vicinity of the 8T and 12T magnetic field regions.   }

\label{fig1}
\end{figure}

It is well-established \cite{DasSarma2005} that the lifting of the electron spin degeneracy by an in-plane magnetic field can lead to an increase in
resistance due to the suppression of the screening of charged
impurities. This phenomenon is however only
important in low electron density systems where a noticeable spin
polarization builds up in the presence of an external magnetic
field. In the simplest picture, the single particle spin polarization of a 2D electron system at $T=0$ is
proportional to $(m^{*}g^{*}B_{\parallel})/n$, where $m^{*}$ is the
electron effective mass and $g^{*}$ the electronic g-factor. In
GaAs where $g^{*}$ and $m^{*}$ are relatively small, the single
particle spin polarization for our sample with an electron density $n\simeq 10^{11} cm^{-2}$
is less than 4\% in a $10T$ magnetic field. This modest spin polarization buildup cannot account for
the substantial CMR effect observed here with magnetic fields as low as $10T$. Furthermore,
the largest possible spin-polarization induced CMR effect is a factor of four \cite{DasSarma2005} in the metallic ($k_{F}\lambda \gg 1$) phase, much weaker
than what we report in this work.

The temperature dependence of the longitudinal resistance was also measured from $0.18K$ to $1.05K$, and is shown in the panel (a) of Fig.2 at several values of the magnetic field. The normalized resistivity is shown in the panel (b) of Fig.2, with the dotted lines being guide-to-the-eye. In this temperature range, the CMR effect manifests very little temperature dependence for fields lower than $5T$ (not shown in Fig.2). Examinations of other data taken below $5T$ exhibit a weak metallic-like temperature dependence with a small coefficient of resistivity $\frac{\partial \rho}{\partial T}>0$.
In contrast to the low-field (or zero-field) absence of any appreciable temperature dependence in our measured resistivity, consistent with the resistivity of high-mobility 2D GaAs systems  \cite{Lilly} with density $n\sim 10^{11}$ $cm^{-2}$, the high-field data ($\gtrsim 10T$) exhibits some complex temperature dependence. The data above $12T$ manifest strong metallic temperature dependence with $\rho(T)$ increasing almost linearly with temperature by $\sim$ 20\% or more for $T\simeq 0.2-1K$. This contrasts greatly with the low-field data (for $B_{\parallel}\lesssim 5T$), where $\rho(T)$ changes by less than a few percent in the same temperature range. In between the low  ($\lesssim 5T$) and high-field ($\gtrsim 12T$) regimes, $\rho(T)$ shows a complex and non-monotonic behaviour, as can be readily seen in Fig.1 (b) and (c), and Fig.2 (b).\\


It is apparent from Fig.1a that the 40 nm sample exhibits two kinks in its resistivity at $B_{\parallel}\sim 8T$ and $11T$, with the kink sharpness  being stronger at lower temperatures (plotted on an expanded scale in Fig.1b and Fig.1c). No such  kink is  observed in the data of the 30 nm sample. The measured resistivity, particularly at the lowest temperature (solid curve at $T=0.18K$),  shows a clear non-monotonicity as a function of the magnetic field, with the non-monotonic feature being more pronounced at 
$\sim$8T, while being suppressed as the temperature is increased. The expanded scale in Fig.1c also indicates that the temperature dependence of the resistivity is non-trivially affected by the kink feature near $\sim$11T, however disappearing quickly above this field.

\begin{figure}[h]
\includegraphics[width=1.05\linewidth,angle=0,clip]{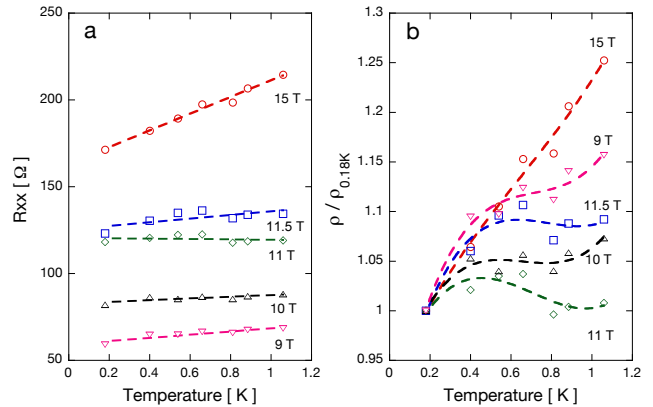}
\caption{A) Longitudinal resistance $R_{xx}$ and B) normalized resistivity versus temperature.
The dotted lines are guides-to-the-eye.  }

\label{fig2}
\end{figure}

We believe that the observed kinks in the magnetoresistance give us a clue about the underlying physics controlling the  CMR phenomenon, indicating that  they arise from the non-perturbative magneto-orbital coupling between the applied in-plane magnetic field and the subband dynamics of the quantum well carriers.  This is most easily seen by comparing the field-induced magnetic length
$l_{{\parallel}}$ with the confinement width of the 2D electron wavefunction $d_{z}\equiv \sqrt{<z^{2}>}$, with the 2D electron layer and $B_{\parallel}$ being in the x-y plane. The condition $d_{z}=l_{\parallel}$ occurs for our 40 nm quantum well at fields $B_{\parallel}\simeq 9T$,  and at a field of $20T$ in the 30 nm sample. We note that this condition for strong magneto-orbital  coupling $l_{\parallel} \lesssim d_{z}$ is much more stringent than the condition  $l_{\parallel} \lesssim d$ , occurring at a
field of only 1T for both the $d=40nm$ and $d=30nm$  quantum well samples. The true quasi-2D transverse width of the quantum well wavefunction in the z-direction is thus much smaller than the physical well width. The regime $l_{\parallel} \ll d_{z}$ involves very strong magneto-orbital coupling between the Landau level and the transverse subband dynamics. In this case, the quasi-2D confinement potential and the applied magnetic field are equally important in controlling the carrier dynamics of the electron system since $\hbar \omega_{c}\gg E_{ij}$, where $\hbar \omega_{c}$ is the in-plane cyclotron energy and  $E_{ij}$ are the subband energy differences associated with the quantum well confinement in the z-direction. In such a situation, realized in our 40 nm quantum well for $B_{\parallel}\gtrsim 8T$, the system can no longer be considered a 2D system because of the non-perturbative magneto-orbital coupling.

To emphasize and reinforce the fact that our observed high parallel field state is a new metallic state in spite of a factor $\sim$300 enhancement in resistivity, we show in Fig. 3 its comparison with data taken in a  tilted magnetic field, where a 40 nm sample cut from the same wafer is now subjected simultaneously to a large parallel and a large perpendicular field. 
Whereas the strict parallel field case produces a  factor of $\sim$300 CMR as shown in Fig. 1, the tilted field case  produces a longitudinal magnetoresistance that is several orders of magnitude larger  at high parallel magnetic field, indicating a clear insulating state.  We have previously identified this tilted field situation as a possible Wigner crystal  \cite{Piot} whereas we identify our currently reported state in the presence of zero perpendicular field as a novel metallic state driven solely by the applied parallel field through magneto-orbital coupling.

Theoretically, calculating the field and temperature-dependence of the magnetoresistance in a quasi- 2D system  with finite thickness,  in the presence of a strong in-plane magnetic field is extremely complex, and beyond the scope of this  experimental work. This complexity arises  when the   in-plane 
magnetic field strength is such that the  magnetic length becomes comparable or shorter than the 2D thickness, $l_{\parallel}\lesssim d_{z}$ and the system becomes quasi-three-dimensional.  All resistive scattering processes (both electron-impurity and electron-phonon) are strongly affected by the in-plane field since the scattering matrix elements depend in this case on the quasi-2D confinement wavefunctions which themselves are  affected non-perturbatively by the external magnetic field.  In other words, one needs to consider a transport theory in the presence of non-perturbative effects of both the confining quantum well potential and the external in-plane magnetic field, a difficult problem that has not yet been solved in any context.  We can, however, crudely estimate the magnitude of the CMR effect by assuming that our extremely high-mobility  sample is limited by resistive scattering from background random (unintentional) charged impurity scattering.
It is then possible to calculate the matrix elements for electron-impurity scattering incorporating the non-perturbative effect of $B_{\parallel}$ in the quantum well wavefunction \cite{Stopa,Stern}, which then immediately leads to an estimate for the CMR effect. We obtain a CMR factor (for the $d=40nm$ sample) of 20 for $B_{\parallel}=16T$ and 200 for $B_{\parallel}=45T$, which are in semi-quantitative agreement with our experimental observations in Fig.1. This qualitatively validates our basic magneto-orbital coupling explanation 
for the CMR effect.

In conclusion, we can make the following concrete remarks about the physics underlying the magneto-orbital CMR effect. 
{\it i)} The temperature dependence for $B_{\parallel}>12T$ likely arises from phonon scattering which is strongly affected by the strong magneto-orbital coupling. This is consistent with the linear temperature dependence of the resistivity for $B_{\parallel}>12T$, shown in Fig.2. {\it ii)} The kink in the resistivity at $B_{\parallel}\simeq 8T$ most likely arises from a magneto-orbital resonance where the first excited magneto-electric subband ({\it i.e.} the quantum well confined levels in the presence of strong magneto-orbital coupling) is pushed down through the Fermi level by the applied field\cite{Stopa}, allowing strong inter-subband scattering in the 2D system.  It is known \cite{DasSarma87} that such inter-subband resonance can produce the kink structure  and the associated temperature dependence observed in this work.   {\it iii)} The kink at $\sim 11.5T$ likely corresponds to the onset of the quasi-3D regime, where the cyclotron energy exceeds  all  confined energy levels. In this regime ($B_{\parallel}>12T)$, the system can be viewed as a quasi-3D electron system with strong scattering from boundaries, impurities and phonons.  {\it iv)}  A 2D-to-3D crossover transition is  expected as the magnetic field is increased  continuously from zero to high magnetic fields. This quasi-three-dimensional dynamics is likely to be important for the CMR observed in this work.

\begin{figure}[tbp]
\vspace*{-10mm}
\includegraphics[width=1.1\linewidth,angle=0,clip]{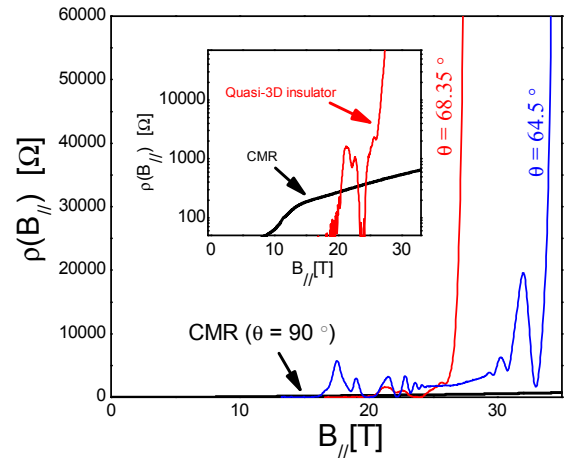}
\vspace*{-16mm}\caption{Comparison of the resistivity versus parallel field  between a tilted 40 nm quantum well \protect\cite{Piot} at $\theta=68.35^{o}$ (red), $\theta=64.5^{o}$ (blue), and the pure parallel field case studied in this work, $\theta=90^{o}$(black). The inset shows the same data  for the $\theta=68.35^{o}$ tilt angle and $\theta=90^{o}$ but on a semi-log scale.    }

\label{fig3}
\end{figure}

This work has been supported by the NSERC, CIFAR, and FQRNT.  
A portion of this work was performed at the National High Magnetic
Field Laboratory, which is supported by NSF Cooperative Agreement
No. DMR-0084173, by the State of Florida, and by the DOE. We also thank T. Murphy, E. Palm, R. Talbot, R. Gagnon and J. Smeros for technical assistance.


\end{document}